\documentclass{emulateapj}
\usepackage{epsfig, graphicx}   

\def\kms    {\ifmmode{{\rm \ts km\ts s}^{-1}}\else{\ts km\ts s$^{-1}$}\fi}
\def\msol   {\ifmmode{{\rm M}_{\odot}}\else{M$_{\odot}$}\fi}
\def\lsun   {\ifmmode{{\rm L}_{\odot}}\else{L$_{\odot}$}\fi}
\def\ts     {\thinspace} 
\def\ci   {\ifmmode{{\rm C}{\rm \small I}}\else{C\ts {\scriptsize I}}\fi}
\def\cone {\ifmmode{{\rm C}{\rm \small I}(1-0)}\else{C\ts {\scriptsize I}(1--0)}\fi}
\def\ctwo {\ifmmode{{\rm C}{\rm \small I}(2-1)}\else{C\ts {\scriptsize I}(2--1)}\fi}
\def\cii  {\ifmmode{{\rm [C}{\rm \small II}]}\else{[C\ts {\scriptsize II}]}\fi}
\def\aco  {\ifmmode{^{12}{\rm CO}(J=1\to0)}\else{$^{12}{\rm CO}(J=1\to0)$}\fi}
\def\bco  {\ifmmode{^{12}{\rm CO}(J=2\to1)}\else{$^{12}{\rm CO}(J=2\to1)$}\fi}
\def\m    {\ifmmode{\mu {\rm m}}\else{$\mu$m}\fi}
\def\cco  {\ifmmode{^{13}{\rm CO}(J=1\to0)}\else{$^{13}{\rm CO}(J=1\to0)$}\fi}
\def\dco  {\ifmmode{^{13}{\rm CO}(J=2\to1)}\else{$^{13}{\rm CO}(J=2\to1)$}\fi}
\def\eco  {\ifmmode{^{12}{\rm CO}(J=3-2)}\else{$^{12}{\rm CO}(J=3-2)$}\fi}
\def\hi   {\ifmmode{{\rm H}{\rm \small I}}\else{H\ts {\scriptsize I}}\fi}
\def\hii  {\ifmmode{{\rm H}{\rm \small II}}\else{H\ts {\scriptsize II}}\fi}
\def\ha   {\ifmmode{{\rm H}{\alpha}}\else{H${\alpha}$}\fi}
\def\hh     {\ifmmode{{\rm H}_2}\else{H$_2$}\fi}
\def\nhh     {\ifmmode{N({\rm H}_2)}\else{$N$(H$_2$)}\fi}
\def\tex {\ifmmode{{T}_{\rm ex}}\else{$T_{\rm ex}$}\fi}
\def\tmb {\ifmmode{{T}_{\rm mb}}\else{$T_{\rm mb}$}\fi}
\def\tkin {\ifmmode{{T}_{\rm kin}}\else{$T_{\rm kin}$}\fi}
\def\microns {\ifmmode{\mu{\rm m}}\else{$\mu$m}\fi}
\def\nhh   {\ifmmode{n({\rm H}_2)}\else{$n$(H$_2$)}\fi}

\shorttitle{Dense Molecular Gas in the Outflow of NGC\,253}
\shortauthors{Walter, Bolatto, Leroy, et al.}

\begin{document}

\title{Dense Molecular Gas Tracers in the Outflow of the Starburst Galaxy NGC\,253}
  
\author{
Fabian Walter\altaffilmark{1,2,3}, 
Alberto D. Bolatto\altaffilmark{4,5}, 
Adam K. Leroy\altaffilmark{6}, 
Sylvain Veilleux\altaffilmark{4}, 
Steven R. Warren\altaffilmark{7,4}, 
Jacqueline Hodge\altaffilmark{8}, 
Rebecca C. Levy\altaffilmark{4},
David S. Meier\altaffilmark{9,2},
Eve C. Ostriker\altaffilmark{10}, 
J\"{u}rgen Ott\altaffilmark{2}, 
Erik Rosolowsky\altaffilmark{11}, 
Nick Scoville\altaffilmark{3},
Axel Weiss\altaffilmark{12},
Laura Zschaechner\altaffilmark{1},
Martin Zwaan\altaffilmark{13}
}

\altaffiltext{1}{Max Planck Institute f\"ur Astronomie, K\"onigstuhl 17, 69117, Heidelberg, Germany}
\altaffiltext{2}{National Radio Astronomy Observatory, PO Box O, 1003 Lopezville Road, Socorro, New Mexico 87801, USA}
\altaffiltext{3}{Astronomy Department, California Institute of Technology, MC105-24, Pasadena, California 91125, USA}
\altaffiltext{4}{Department of Astronomy, Laboratory for Millimeter-wave Astronomy, and Joint Space Institute, University of Maryland, College Park, Maryland 20742, USA}
\altaffiltext{5}{Visiting, Max-Planck Institute for Astronomy, Heidelberg, Germany}
\altaffiltext{6}{Department of Astronomy, Ohio State University, 100 W 18th Ave, Columbus, OH 43210, USA}
\altaffiltext{7}{Cray, Inc., 380 Jackson Street, Suite 210, St. Paul, MN 55101, USA}
\altaffiltext{8}{Leiden Observatory, Niels Bohrweg 2, 2333 CA Leiden, Netherlands}
\altaffiltext{9}{New Mexico Institute of Mining \& Technology, 801 Leroy Place, Socorro, NM 87801, USA}
\altaffiltext{10}{Department of Astrophysical Sciences, Princeton University, Princeton, New Jersey 08544, USA}
\altaffiltext{11}{Department of Physics, University of Alberta, Edmonton, AB, Canada}
\altaffiltext{12}{Max-Planck-Institut f\"ur Radioastronomie, Auf dem H\"ugel 69, Bonn, Germany}
\altaffiltext{13}{European Southern Observatory, Karl-Schwarzschild-Strasse 2, 85748 Garching, Germany}

\begin{abstract} 

We present a detailed study of a molecular outflow feature in the
nearby starburst galaxy NGC\,253 using ALMA. We find that this feature
is clearly associated with the edge of NGC\,253's prominent ionized
outflow, has a projected length of $\sim$300\,pc, with a width of
$\sim50$\,pc and a velocity dispersion of $\sim40$\,km\,s$^{-1}$,
consistent with an ejection from the disk about 1\,Myr ago.  The
kinematics of the molecular gas in this feature can be interpreted
(albeit not uniquely) as accelerating at a rate of
1\,km\,s$^{-1}$\,pc$^{-1}$. In this scenario, the gas is approaching 
escape velocity at the last measured point. Strikingly, bright tracers of dense molecular
gas (HCN, CN, HCO$^+$, CS) are also detected in the molecular outflow:
We measure an HCN(1--0)/CO(1--0) line ratio of $\sim1/10$ in the
outflow, similar to that in the central starburst region of NGC\,253
and other starburst galaxies. By contrast, the HCN/CO line ratio in
the NGC\,253 disk is significantly lower ($\sim1/30$), similar to
other nearby galaxy disks. This strongly suggests that the streamer
gas originates from the starburst, and that its physical state does
not change significantly over timescales of $\sim1$\,Myr during its
entrainment in the outflow. Simple calculations indicate
that radiation pressure is not the main mechanism for driving the outflow.
The presence of such dense material in
molecular outflows needs to be accounted for in simulations of
galactic outflows.

\end{abstract}


\section{Introduction}

Galactic--scale winds are an ubiquitous phenomenon in both starburst
galaxies and galaxies that host active galactic nuclei (e.g., Veilleux
et al.\ 2013). They are thought to be especially important at high
redshift, where strongly starforming galaxies on the so--called galaxy
`main--sequence' dominate the star formation budget, and the number
density of active galactic nuclei is greater.  Winds provide negative
mechanical feedback, and have been invoked to resolve a number of
important issues in cosmology and galaxy evolution (see e.g., Veilleux
et al.\ 2005 for a review). Although galactic winds have been detected
in galaxies out to high redshift, only in the nearby universe can they
be observed panchromatically in emission to understand the mechanisms
responsible for ejecting the gas. Steadily expanding observations and
simulations show that winds include cold (neutral atomic and
molecular) gas components, both in starburst--driven as well as 
AGN--driven outflows (e.g., in the case of M\,82: Walter et al.\
2002, Engelbracht et al.\ 2006, Roussel et al.\ 2010, Leroy et al.\
2015a, other galaxies: e.g., Rupke et al.\ 2002, Alatalo et al.\ 2011, Aalto et al.\ 2012,
Meier \& Turner 2012, Rupke \& Veilleux 2013), possibly through the entrainment of ambient
material. This cold gas is very difficult to measure, but likely
constitutes the mass--dominant phase in galactic winds. Mechanisms
that have been proposed to drive cool winds include direct radiation
forces (e.g., Murray et al.\ 2011), cosmic ray pressure gradients
(e.g., Uhlig et al.\ 2012), pressure due to supernovae-driven
superbubbles (Fujita et al.\ 2009, Bolatto et
al.\ 2013), and progressive entrainment into the ionized flow
facilitated by Kelvin--Helmholtz instabilities (Heckman et al.\
2000). Recent models suggest that the cold gas could also be emerging
from condensations of the hot ionized gas phase through runaway
thermal instabilities (e.g., Faucher-Gigu{\`e}re \& Quataert 2012,
Zubovas \& King 2012, Nayakshin \& Zubovas 2012, Thompson et al.\
2016, Bustard et al.\ 2016). The latter predictions imply a small
spatial offset between the hot burst and the emergence of cold
material condensing out of the wind. In most cases, observational 
constraints on the velocity of the outflowing gas, both neutral and ionized,
indicate that the gas does not reach the escape velocity (above references), 
i.e. that the currently outflowing gas may be re--accreted at later cosmic times.

NGC\,253 is one of the best laboratories to study starburst--driven
galactic--scale winds in detail due to its proximity
($D$\,=\,3.5\,Mpc, Rekola et al.\ 2005). It is known for the galactic
wind emerging from its central $\sim$200\,pc (e.g., Sharp \&
Bland--Hawthorn 2010). The wind is seen in H$\alpha$ and X--ray
emission (Strickland et al.\ 2000, 2002, Westmoquette et al.\ 2011), in
neutral gas (Heckman et al.\ 2000), warm H$_2$ (Veilleux et al.\ 2009),
and in OH emission and absorption (Turner 1985, Sturm et
al. 2011). HST imaging reveals the entrained dust emission in optical
broad--band imaging, already suggesting that the wind also carries
significant amounts of molecular gas.

ALMA cycle~0 imaging revealed the molecular wind in NGC\,253 in
CO(1--0) emission (Bolatto et al.\ 2013). This molecular wind carries
enough mass to substantially shorten the current star formation
episode in this galaxy, or to definitively quench star formation over
a much longer period if a substantial fraction of the gas were to
reach escape velocity. In the interferometer imaging, the wind breaks
up in molecular filaments or ``streamers'' emerging from the
central starburst area. The brightest of these streamers is West of
the central starburst and points to the South outlining an edge of the
approaching side of the approximately conical ionized outflow
(Westmoquette et al.\ 2011); we hereafter refer to it as the
(south--west) {\em SW streamer}.  In Meier et al.\ (2015) we present a
cartoon of the central starburst region of NGC\,253, and discuss other
tracers of the molecular gas found in the central starburst region of
NGC\,253, while in Leroy et al.\ (2015b) we focus on the properties of
the giant molecular clouds and the structure of the starburst itself.

In this study we present and discuss the properties of this brightest
outflowing molecular gas streamer, as obtained from the combined
cycle~0 and~1 ALMA observations together with new IRAM single-dish
data. This paper is structured as follows: \S2 summarizes the
interferometric, single--dish, and Hubble Space Telescope  observations, and \S3
shows the results. In \S4 we present a discussion of our findings, and
\S5 summarizes our conclusions.  We assume
a distance to NGC\,253 of $D=3.5$\,Mpc (Rekola et al.\ 2005), leading
to a linear (projected) scale of 1\arcsec\,=\,17\,pc.

\begin{figure*} [t]
\centering
\includegraphics[width=9.0cm,angle=0]{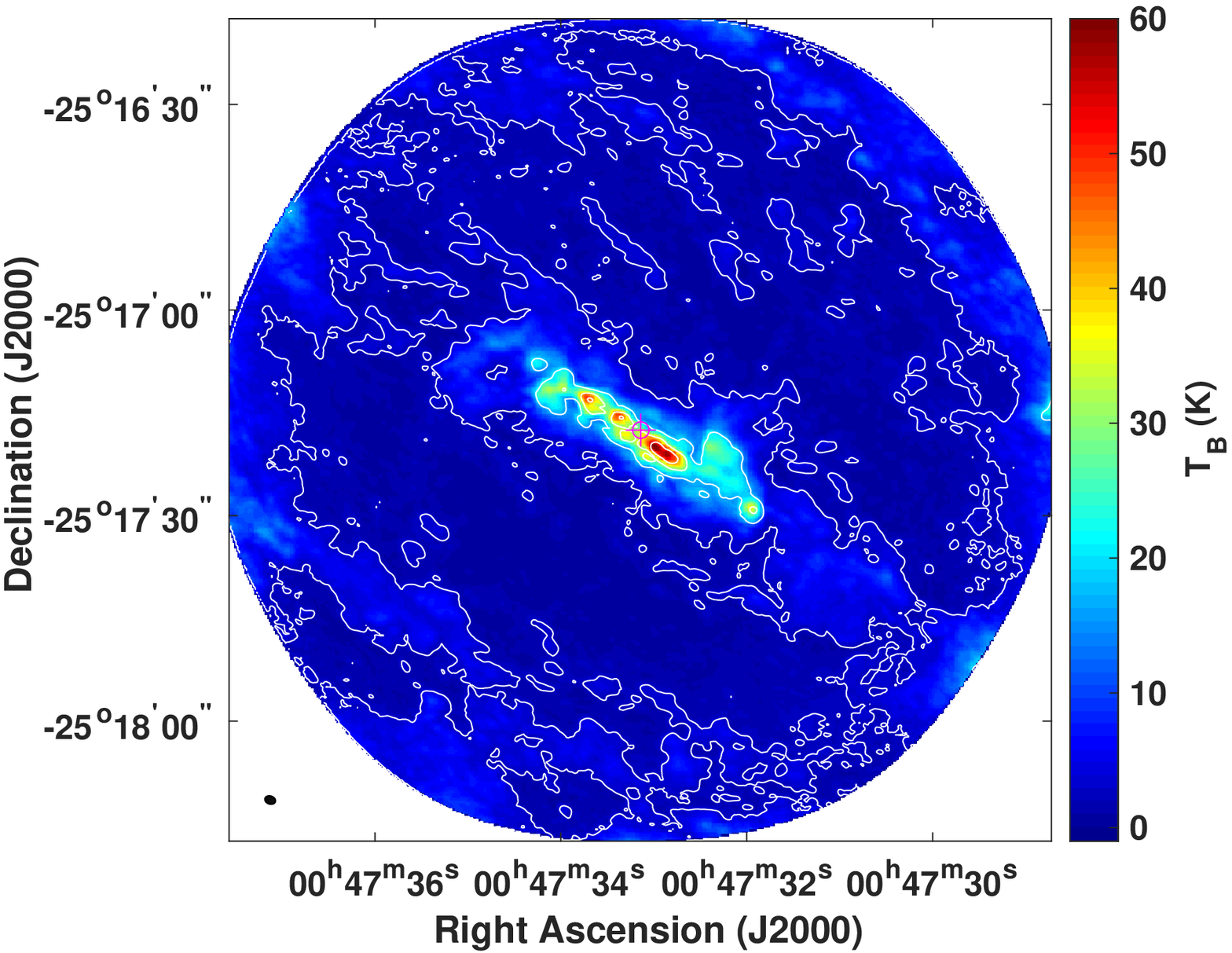}\includegraphics[width=9.0cm,angle=0]{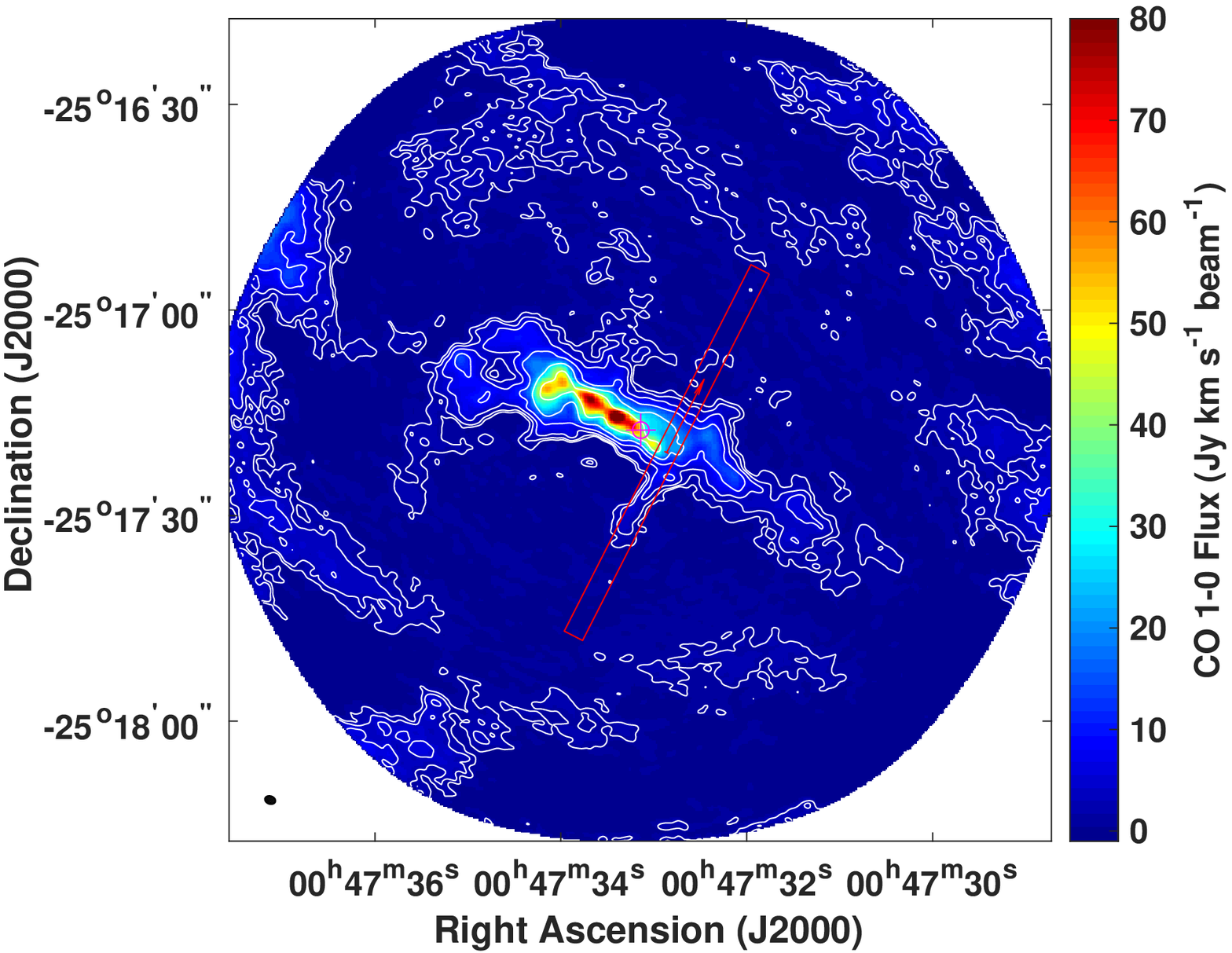}
\caption{ALMA CO(1--0) observations of NGC\,253. The kinematic center determined by  M\"uller-S\'anchez et al.\ (2010) at $00^{\rm h}47^{\rm m}33\fs14$, $-25^\circ17\arcmin17\farcs52$ (J2000) is indicated with a magenta cross, with the $\pm1.2\arcsec$ $3\sigma$ uncertainty shown by the magenta circle. {\bf Left:} peak brightness map of the CO(1--0) emission in
  NGC\,253 covered by our 7--point ALMA band~3 mosaic (we present here
  the Briggs-weighted observations with a robust parameter of 0.5,
  which have a synthesized beam of $1.6\arcsec\times1.2\arcsec$ shown
  in the left bottom corner). Contours are drawn at 2.5, 5, 10, 20,
  30, 40, and 50~K. The peak surface brightness in the map is
  60~K. The emission is dominated by the bright central bar, but
  emission from the spiral arms in the disk is also visible towards
  the SE and NW edges of the covered area. Note that this map is
  corrected for the primary beam attenuation, so the noise increases
  at the edges of the mosaic. {\bf Right:} velocity--integrated CO
  (1--0) flux density $S_\nu\Delta v$ over the velocity range
  120--255~km\,s$^{-1}$, highlighting the {\em SW streamer} of the
  outflow. The contours correspond to 1.25, 2.5, 5, 10, 20, 40
  Jy\,beam$^{-1}$\,km\,s$^{-1}$.  The red rectangle shows the
  orientation of the position-velocity diagram shown in
  Figs.~\ref{fig:fig2}--\ref{fig:fig3}, and
  \ref{fig:fig5}--\ref{fig:fig6} (centered at $00^{\rm h}47^{\rm
    m}32\fs86$, $-25^\circ17\arcmin20\farcs8$ with a position angle of
  333$^\circ$ so that offsets are positive N of the central bar). The
  heliocentric velocity of the central bar material in this
  position-velocity cut is 280\,km\,s$^{-1}$ (see for example the top
  panel of Fig. \ref{fig:fig6}).
\label{fig:fig1}} 
\end{figure*} 

\begin{figure} 
\begin{center}
\includegraphics[width=\columnwidth,angle=0]{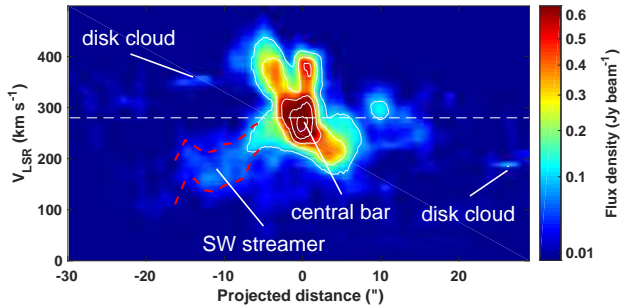} 
\caption{Position--velocity CO(1--0) diagram along the slit shown in
  Fig.~\ref{fig:fig1} (equivalent to a long-slit spectrum). The emission is
  dominated by the intense emission from the central starburst
  region. The narrow features at offsets $\sim\pm20"$ are due to
  molecular clouds in the galaxy's rotating disk (labeled disk
  cloud). The CO (1--0) outflow {\em SW streamer} (negative offsets)
  is also labeled. The white dashed line shows the adopted reference
  velocity for emission in the central bar at this location in the
  galaxy, of 280\,km\,s$^{-1}$. The contours indicate S$_\nu$ of 0.09,
  0.25, 0.49, 0.81, and 1.21 Jy\,beam$^{-1}$. The color scale uses 
  a square root stretch, and the noise in the data is 
1.6\,mJy\,beam$^{-1}$ in emission-free velocity ranges. A typical brightness for
streamer material between the dashed lines is 40--60\,mJy\,beam$^{-1}$} . \label{fig:fig2}
\end{center}
\end{figure}

\section{Observations}

\subsection{ALMA Observations}

The data presented here are based on ALMA cycle 0 and 1 observations
and include IRAM 30\,m and Mopra single dish observations to account
for the missing short spacings in the interferometric ALMA
imaging. 

Two frequency setups were observed in ALMA band 3, one
high--frequency setup (LSB: 99.8--103.7 GHz; USB: 111.8--115.7 GHz),
that covers the (redshifted) CO(1--0) and CN lines, and one low--frequency setup
(LSB: 85.6--89.6 GHz; USB: 97.4--101.4 GHz) that covers key
high--density tracer molecules,  such as HCN, HNC and HCO$^+$
(see Meier et al.\ 2015 for a full overview of all transitions covered
by the observations).
Additional details about the ALMA observations and reduction can be
found in Bolatto et al.\ (2013), Leroy et al.\ (2015b), and Meier et
al. (2015). The ALMA band 3 interferometric observations include:

{\em ALMA high--frequency setup:} A 7--point mosaic targeting the
CO(1--0) line to map the central 1\,kpc ($\sim$1\arcmin) of NGC\,253
in the CO(1--0) transition These observations used the following
calibrators: J0038-2459 (phase), J2357-5311 (bandpass) and Uranus
(amplitude). Cycle 1 observations were obtained on 19-Nov-2013,
01-Dec-2013, and (twice) on 02-Dec-2013. The total on--source
integration time was 2.5\,h, using typically 36 antennas. These
observations were complemented by a 2.6\,h, 7-pointing mosaic with the
ALMA compact array (ACA) covering $\sim$2$'\times$2$'$
(2\,$\times$\,2\,kpc) at 75\,pc resolution (cycle 1). These ACA
observations used the following calibrators: J0038-2459 (phase),
J2258-2758 (bandpass) and Neptune (amplitude). The ACA observations
were executed on 07-Oct-2013, twice on 01-Nov-2013, three times on
05-Nov-2013, three times on 06-Nov-2013, and on 14-Nov-2013. The ACA
observations were obtained using 9 antennas.

{\em ALMA Low--frequency setup:} A 3--pointing mosaic at 90\,GHz along
the NGC\,253 bar (cycle 1), targeting the high-dipole molecules
CS(2--1), HCN(1--0), HCO$^{+}$(1--0), and a host of other molecules
contained in the band (Meier et al.\ 2015). These observations used the
following calibrators: J0038-2459 (phase), J2258-2758 (bandpass) and
Neptune (amplitude). Observations were done on 31-Aug-2014, with a
total of 27.5 minutes on--source using 32 antennas.

The cycle~1 data were combined with the cycle~0 data published in
Bolatto et al.\ (2013) and Meier et al.\ (2015), and the combined data
were used in the analysis by Leroy et al.\ (2015).  Because the
continuum in the NGC\,253 starburst is very bright it is possible to
self-calibrate on it. We applied self-calibration to remove residual
phase and flux calibration errors inherent to the data and attain a
dynamic range higher than otherwise possible. Phase information using
the continuum were self-calibrated first followed by 
amplitude self-calibration.

Natural and Briggs robust--weighted, continuum--subtracted, cubes were
created at 5\,km\,s$^{-1}$ velocity resolution. All values were
primary beam corrected for all quantitative analysis. As a final step,
for all comparisons we convolve the low-- and high--frequency setups
to a common beam size of 1.9\arcsec\ (32\,pc).  The final robust-weighed 
CO(1--0) cube has an rms of 1.6\,mJy\,beam$^{-1}$ in 5 km\,s$^{-1}$ channels 
and a beam size of 1.6$"\times$1.2$"$ with PA of 71.5$^\circ$. In velocity ranges where
signal is present, deconvolution artefacts dominate the noise (despite self-calibration and 
careful cleaning the data is dynamic--range limited), and the practical noise floor is closer to 6.5\,mJy\,beam$^{-1}$.
For the HCN, CN, CS, and HCO$^+$ lines, the rms is 1 mJy\,beam$^{-1}$
for a 1.9$"\times$1.9$"$ beam.

\begin{figure} \centering
\includegraphics[width=\columnwidth,angle=0]{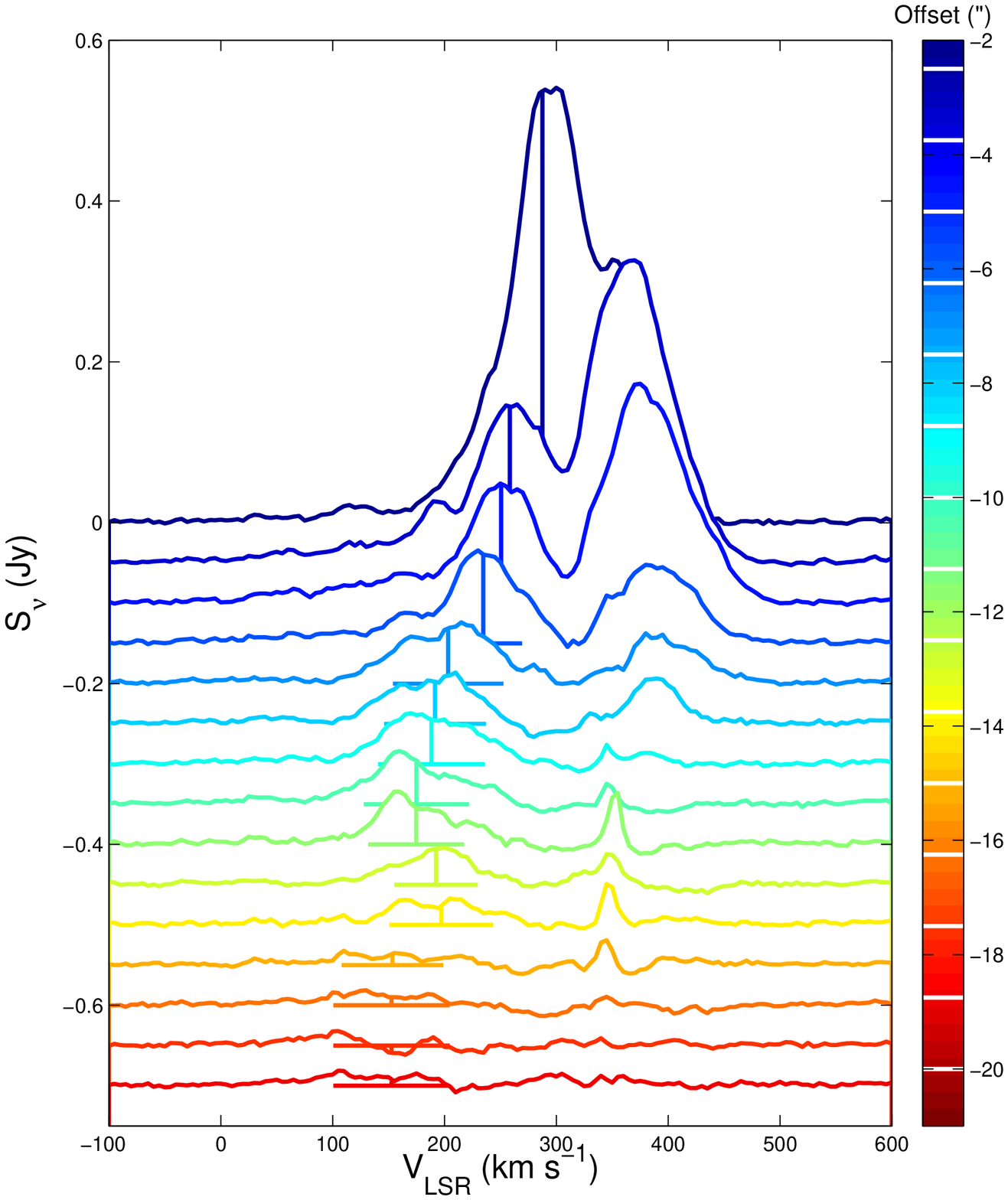} 
\caption{CO(1--0) spectra along the southern part of the
  position-velocity cut in Figs. \ref{fig:fig1} and \ref{fig:fig2}, following the {\em SW
    streamer}. The colors indicate offsets from the plane, as
  indicated by the color bar to the right (spectra are offset
  vertically by 0.05\,Jy for display purposes). At small offsets, the
  emission is dominated by the starburst in the central bar, and for
  intermediate offsets (--4$"$to --10$"$) line splitting is
  observed. The diffuse molecular outflow is then picked up at v$_{\rm
    lsr}\sim$200\,km\,s$^{-1}$, and can be traced out to offsets of
  --18$\arcsec$ (the horizontal and vertical bars indicate the
  Gaussian--fit FWHM, central velocity, and amplitudes in the
  streamer). Given the low S/N of the three spectra at 
 the largest offsets, we do not use their fits in our subsequent 
 analysis}. The narrow CO(1--0) feature around v$_{\rm
    lsr}\sim$350\,km\,s$^{-1}$ is due to molecular gas emission from a
  GMC in the disk of NGC\,253, labeled in Fig. \ref{fig:fig2} as
  ``disk cloud''. \label{fig:fig3new}
\end{figure}

\subsection{Single Dish Observations}

In the absence of ALMA total power measurements, we have corrected the
ALMA interferometric imaging for zero spacings using the following
single dish telescopes:

{\em Mopra:} We use measurements obtained by the Mopra telescope to
correct the ALMA CO(1--0) observations (high--frequency setup), as
discussed in in Bolatto et al.\ (2013), Meier et al.\ (2015), and
Leroy et al.\ (2015). 

{\em IRAM 30\,m:} We use new IRAM 30\,m telescope measurements to
correct the high--density tracer molecules (low--frequency
setup). These observations cover all relevant tracer molecules,
e.g., HCN, HCO$^+$, CS (project 209--14 during
4--9-Mar-2015). NGC\,253 was above
20$^{\circ}$ for $\sim$4.5 hours each day, resulting in a total of
12.6\,h of on--source time with exceptional observing conditions.  To
map a $2\arcmin\times2\arcmin$ field around the center of NGC\,253
with uniform sensitivity, we obtained 36 back and forth on-the-fly scan--maps with
4\arcsec\ spacing, alternating both right ascension and
declination. We used the E090 HIGHDENS receiver setup spanning 81.335 to 89.335 GHz
in the LSB and 97.335 to 105.335 GHz in the USB, allowing for simultaneous
observations of high-density tracers including HCO$^+$, HCN, and CS.
We observed Mars for pointing, focus, and calibration.  The map has a
sensitivity of 4\,mK in a 3\,km\,s$^{-1}$ channel at 89\,GHz.  The
frequency-dependent beam size ranges from 26\arcsec\ to 29\arcsec.

In the case of CO(1--0), we combined the ALMA cube with the Mopra
single dish data using the feathering Miriad task {\tt immerge} with a
gain of 16.2 Jy/K. The resulting cube was used as the model input in
the CASA {\tt clean} task on the ALMA--only data. We combined
zero-spacings for the high--density tracer observations from the IRAM
30\,m telescope with the ALMA cubes using the CASA task {\tt feather}.
For the ALMA--IRAM\,30m combination, prior to feathering, we applied a
gain factor of 6.15 Jy/K on the 30-meter cubes. As in the CO($1-0$)
combination, the final cubes were created using the feathered cubes as
models for the inversion of the ALMA data.  This was done by including
the feathered cubes as models in the ALMA dataset using {\tt setjy},
followed by {\tt clean} using the same parameters as those used to
create the original ALMA cubes. In both cases this zero spacing
correction significantly improved the negative bowls in the original
(ALMA 12m--only) data cubes.

\begin{figure*} \centering
\includegraphics[width=18.0cm,angle=0]{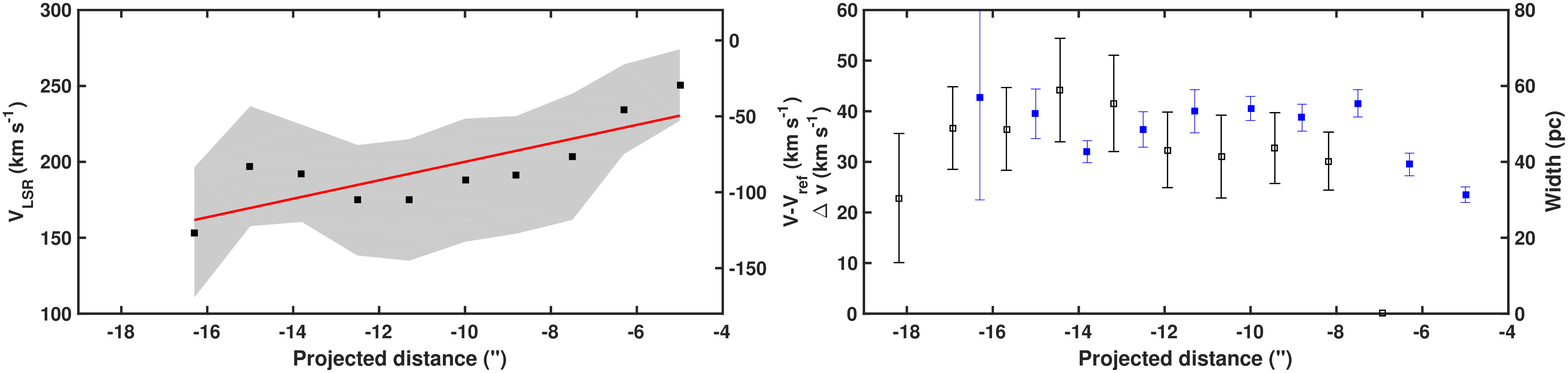} 
\caption{Fit results for the CO emission in the {\em SW
    streamer}. {\bf Left:} Line-of-sight velocity versus projected
  distance. Note that we estimate that the projection correction is a
  factor of $\sim3$ for velocity (see text), while it is almost
  negligible for distance. The points show the central velocity as a
  function of position along the slit (origin is the center of the
  slit which corresponds to the center of the bright bar emission;
  c.f., Fig. \ref{fig:fig1}). The grey region indicates the FWHM of
  the CO line at the different positions. The red line is a linear fit
  to the central velocity data. The left y--axis labels indicate the
  measured velocities, whereas the velocities on the right y--axis are
  shown relative to a fiducial reference spectrum at $0$ offset
  ($V_{ref}=280$ km\,s$^{-1}$; Fig. \ref{fig:fig2}). {\bf Right:}
  Velocity and spatial width versus projected distance: The black open
  squares show the FWHM of the spectral line (indicated as a grey
  region in the left panels) as a function of offset, with
  corresponding scale on the left axis. The blue points show the
  spatial FWHM of the {\em SW streamer} (deconvolved), measured using
  a moment map created for the relevant channels, with corresponding
  scale on the right axis.\label{fig:fig3}}
\end{figure*}

\subsection{Hubble Space Telescope Observations}

We compare our ALMA observations to the ionized gas outflow, traced by
{\em Hubble Space Telescope} imaging of hydrogen recombination
lines. We use a WFPC2 H$\alpha$ image from Proposal ID 5211, first
presented in Watson et al.\ (1996). That project observed NGC\,253 in
the F656N (on) and F675W (off) filters. We pair this with
Paschen--$\beta$ imaging from two projects (Proposal IDs 12206 and
13730), which observed the nuclear region in the F128N (on) and F130N
(off) filters. In both cases we use the {\em Hubble} legacy archive
enhanced data products, beginning with the drizzled images. For the
H$\alpha$ we found it necessary to combine the two visits to reject
cosmic ray artifacts. Then for both sets of images, we aligned the
on-- and off--line filters and fit a median scaling between the two in
a region near but not in the nucleus. The scaling factor derived was
0.0241 and 0.95 for F675W and F130N, respectively. We used this
scaling to subtract the stellar contamination from the narrow-band
on-line image. We then converted the resulting continuum-subtracted
image to have units of erg~s$^{-1}$~cm$^{-2}$~sr$^{-1}$ using the
observatory supplied counts--to--flux conversion ({\tt PHOTFLAM}
keyword) and bandpass width ({\tt PHOTBW} keyword). The values
employed for H$\alpha$ and Paschen--$\beta$ were $1.461\times10^{-16}$
and $4.278\times10^{-19}$ for {\tt PHOTFLAM}, and 53.768 and 357.438
for {\tt PHOTBW}, respectively. Finally, both images were convolved to
$1.5\arcsec$ resolution and aligned to the ALMA astrometric grid.

\section{Results \& Discussion}

\subsection{CO(1--0) map of the full FOV}

For reference we show the CO(1--0) peak brightness map of our ALMA
mosaic in Fig.~\ref{fig:fig1}. The emission is dominated by the central bar in
NGC\,253 that hosts the nuclear starburst. The second panel in Fig.~\ref{fig:fig1}
shows the CO(1--0) emission integrated over a velocity range that
highlights the {\em SW streamer}.  The position and
orientation of the position--velocity cut used in our subsequent
analysis of the {\em SW streamer} is illustrated by the red box.

\subsection{Spectra Along the SW Streamer}

The position--velocity ($pv$) diagram of the {\em SW streamer}  is shown in Fig.~\ref{fig:fig2},
where the prominent outflow is highlighted with dashed red lines;
these lines represent the FWHM resulting from
Gaussian fitting as a function of position offset. As the orientation of this
$pv$ diagram is essentially perpendicular to the major axis, we do not expect the velocities to
be affected by the rotation of the disk. A different view of
the same data is shown in Fig.~\ref{fig:fig3new}, which plots the
CO(1--0) spectra along the $pv$ diagram for offsets. 
The spectral features identified with the
streamer and the results from the Gaussian fits (central velocity,
peak, and velocity width) are also displayed in Fig.~3.

The Gaussian fitting results are summarized in
Fig.~\ref{fig:fig3}. The left panel shows the central velocity and
FWHM velocity width ($\Delta v$) of the emission as a function of
position along the $pv$ diagram. A first order polynomial fit to the
central velocity as a function of position is overplotted as a red
line.  The right panel shows the velocity and deconvolved (by
subtracting the beam size in quadrature) FWHM spatial width as a
function of position: the outflow has a deconvolved FWHM width of
approximately 40--50\,pc.  The reference velocity is indicated by a
dashed white line in Fig.~\ref{fig:fig2}. It corresponds to the
approximate gas velocity of the bar at the base of the outflow.

\begin{figure*}\centering
\includegraphics[width=9.0cm,angle=0]{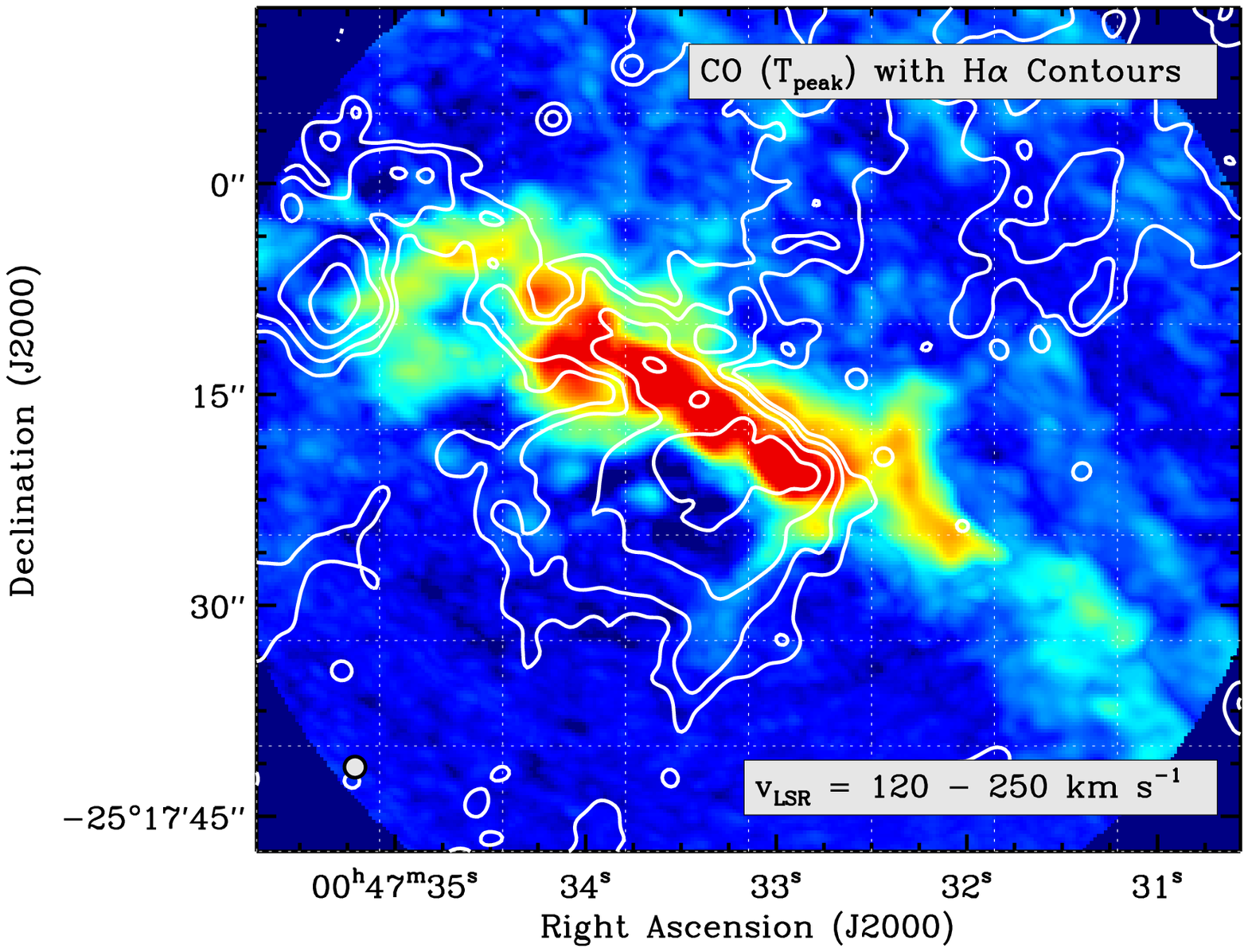}\includegraphics[width=9.0cm,angle=0]{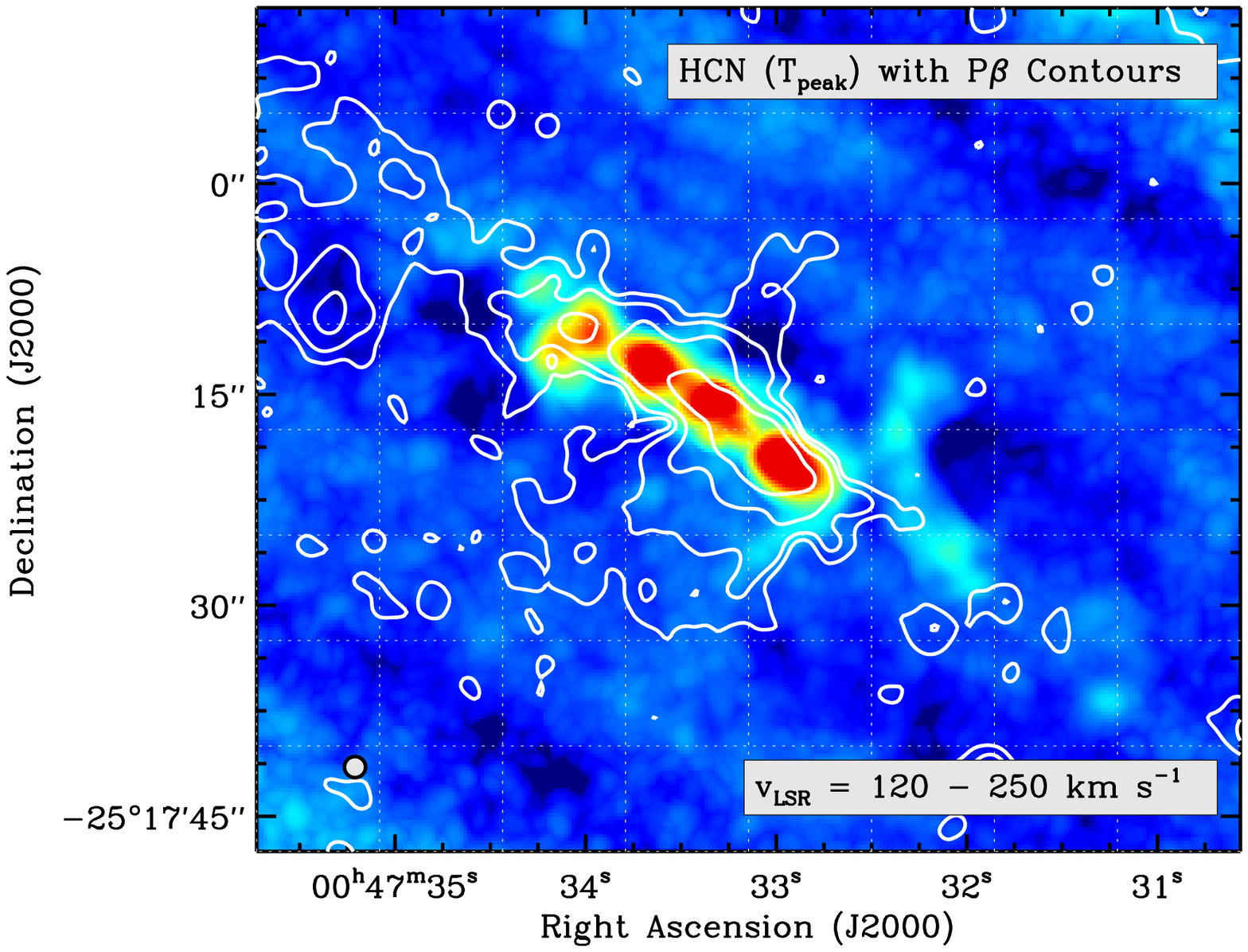}
\caption{Comparison of the molecular gas (ALMA) and ionized gas (HST)
  emission in the outflow. The molecular maps show the peak
  temperature in the emission in CO(1--0) and HCN(1-0) over a velocity
  range of 120-250\,km\,s$^{-1}$, chosen to highlight the {\em SW
    streamer} of the molecular outflow. The ionized gas outflow is
  shown in H$\alpha$ white contours in the left, and in
  Paschen--$\beta$ white contours in the right panel (contours are 1,
  4, 16, $64\times10^{-5}$ erg\,s$^{-1}$\,cm$^{-2}$\,sr$^{-1}$; both
  panels show continuum-subtracted HST observations). The coordinate
  grid is overlaid to facilitate comparison. The {\em SW streamer} is
  already clearly apparent in HCN emission.\label{fig:fig4}}
\end{figure*}

\subsection{Is the Molecular Outflow Accelerating?}

The {\em SW streamer} is blue--shifted with respect to the reference
velocity, implying that the material is approaching us. This blueshift
gets larger with increasing distance from the galaxy plane, which can
be interpreted as an accelerating continuous outflow. This
interpretation of the data is, however, not unique. If the original
ejection occurred over a short period of time and had a distribution
of velocities, the fast ejecta would have gotten farther away than the
slower ejecta, giving rise to a velocity gradient along the streamer.
The apparent projected velocity gradient could also be due to a
gradient in the inclination of the outflow cone, with the (more or
less) constant velocity of CO gas becoming increasingly aligned with
the line of sight at farther distances from the central bar. Such
geometry has been reported in the ionized gas of the bubble of
NGC\,3079 (e.g., Veilleux et al.\ 1994, Cecil et al.\ 2001). The
ionized outflow of NGC\,253, however, seems more like an open--ended
cone than a bubble (Westmoquette et al.\ 2011), suggesting that this
explanation is probably less likely.

The translation of observed line-of-sight velocities into actual
velocities is crucial to determine the mass outflow rate and whether
the material escapes the galaxy or not. The inclination of the galaxy
is 78$^\circ$ (Westmoquette et al.\ 2011), so the projected edge of
the southern outflow cone will have an inclination of
$78^\circ-90^\circ=-12^\circ$, where the negative sign indicates it is
pointing towards us, resulting in blue--shifted emission. If the
emission originates precisely from this projected edge, the $\sin i$
correction would be very large (a factor of $\sim5$). Based on the
median of the distribution of $\tan i$ for the surface of a cone with
the opening angle and orientation as determined by Westmoquette et
al. (2011) for the ionized gas outflow, Bolatto et al.\ (2013) argue
that a correction factor of $\sim3$ is more likely, and we adopt that
value here. Note that, in practice, the actual factor can range from
$\sim1$ to $\infty$ so the projection correction has large
uncertainties.

We measure a gradient of d$v$/d$r=6.1$\,km\,s$^{-1}$/[\arcsec] or
$\sim36$\,km\,s$^{-1}$/[100\,pc] from the polynomial fitting shown in
Fig.~\ref{fig:fig3} (the symmetric ordinary-least-squares bisector yields a 25\%
higher value). Adopting the aforementioned projection correction
factor of 3 which includes both the corrections for distances and
velocities, we derive an actual velocity gradient of
$\sim$\,100\,km\,s$^{-1}$/[100\,pc], or
$\sim$1\,km\,s$^{-1}$/[pc]. Towards the end of the detected outflow we
measure a line--of--sight outflow speed, relative to our reference disk
velocity, of 120\,km\,s$^{-1}$. With the projection correction, this
translates into an actual outflow speed of $\sim$360\,km\,s$^{-1}$.
The escape velocity of NGC\,253 can be approximated using 
equation 16 in Rupke, Veilleux, \& Sanders 2002, and the circular velocity by 
Hlavacek-Larrondo et al.\ 2011. This yields an escape velocity of
 $\sim$500\,km\,s$^{-1}$ (with significant error bars) , i.e. the 
measured outflow velocity is approaching the escape velocity.

The velocity dispersion and width of the {\em SW streamer} are
consistent with the dynamical age of the feature. The dynamical age of
the streamer is $\sim1$\,Myr, corresponding to the approximate ratio
of its deprojected speed of $\sim$300\,km\,s$^{-1}$ and length of
$\sim300$\,pc. The measured velocity dispersion $\Delta
v\sim35-40$\,km\,s$^{-1}$, which would naturally produce a structure
with a width of $\sim40$\,pc in 1\,Myr, very similar to the measured
physical width (Fig.~\ref{fig:fig3}). If the feature were significantly older than
1\,Myr, it would have to be confined (by external pressure or magnetic
fields) to stay as narrow in spatial extent as observed. This implies
that, unless such confinement is present, the outflow velocities
cannot be much slower than 300\,km\,s$^{-1}$.

\subsection{Deep Imaging of the Ionized Gas}

Figure~5 compares the outflow observed in molecular gas (left: CO,
right: HCN) to the well--known ionized gas outflow (left: H$\alpha$, 
right: Pa$\beta$). The tracers of the ionized outflow shown here 
(H$\alpha$ and Pa$\beta$), are significantly affected by extinction
towards the Northern part of the galaxy. The ALMA images correspond to 
peak intensity over the range
$v = 120-250$~km~s$^{-1}$, highlighting the disk, eastern shell, and
western outflow. The {\em SW streamer} aligns closely with
the edge of the outflow as seen in both H$\alpha$ and the more
extinction--robust Pa$\beta$ imaging (contours). In fact the streamer aligns with
bright features in both recombination lines, further solidifying the
association of this feature with the ionized gas outflow. The higher
density of gas near the cold streamer presumably leads to a higher
emission measure and corresponding emissivity of ionized gas. We 
note that the feature seen in dense gas emission on the other side 
of the disk (i.e. opposite to
the {\em SW streamer}) is likely due to peculiar motions in the NGC\,253's 
disk. This region was already identified in early high-density tracer observations 
using OVRO (Knudsen et al. 2007).

The molecular component seems to confine the ionized features in the
left panel of Fig.~\ref{fig:fig4} (CO~+~H$\alpha$), with the ionized gas tapering
to emerge from the center of the circumnuclear starburst. As noted
previously, both recombination lines are brighter to the south of the
galaxy, consistent with that being the approaching side of the
outflow.  Strikingly, extinction due to the surrounding disk is
strong enough that aside from one bright pillar, even Pa$\beta$
emission is not visible north of the starburst region.

Perhaps surprisingly, there is little recombination line emission (and
thus little sign of very recent massive star formation) in the far
western part of the disk. Also, we do not see strong emission centered
on the putative expanding molecular shells identified in Sakamoto et
al.\ (2006). Bolatto et al.\ (2013) argued that these shells are
associated with the molecular streamers in the outflow and are
presumably driven by young stellar clusters of
$M_*\sim10^5$\,M$_\odot$.  If the shells are driven by stellar
feedback, the population driving them must be either older than $\sim
5$~Myr or so embedded as to remain inconspicuous in the Pa$\beta$ image.

\subsection{Dense Gas in the Outflow}

\begin{figure*} \centering
\includegraphics[width=15.0cm,angle=0]{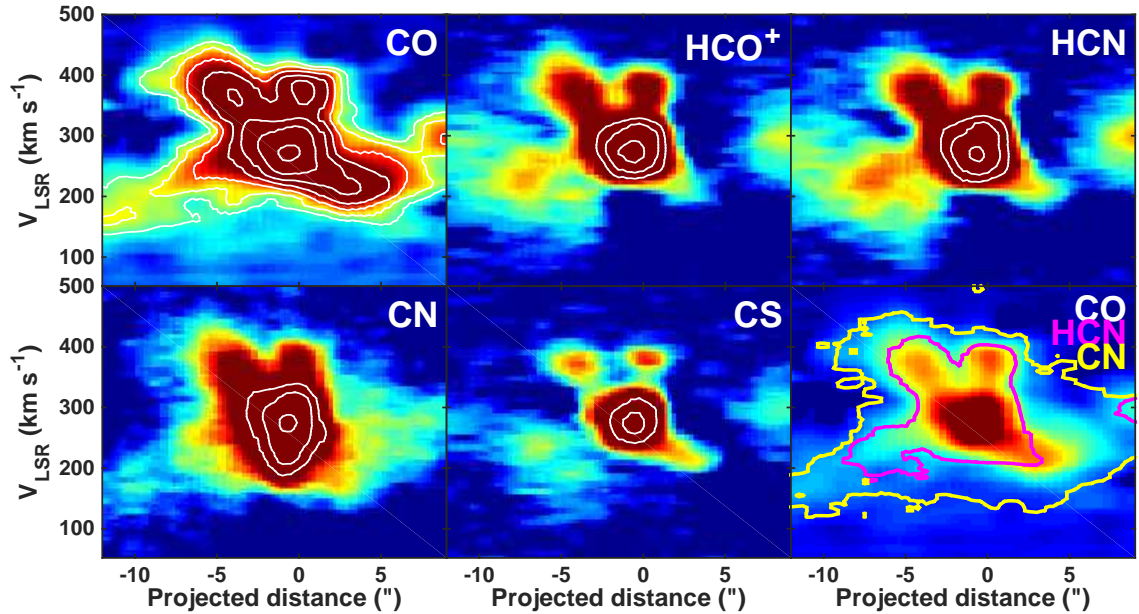} 
\caption{Position--velocity diagrams for a range of gas tracers,
  showing dense-gas emission in the {\em SW streamer} at offsets -5\arcsec to
  -12\arcsec\ (see Fig.~\ref{fig:fig2}).  The panels show emission for individual
  lines convolved to a common resolution of
  1.9\arcsec\ (left-to-right, top-to-bottom): CO(1--0),
  HCO$^{+}$(1--0), HCN(1--0), CN (including blended hyperfine
  structure, see \S4.1.3), and CS(1--0). The white contours are at 50,
  90, 160, 290, 530, 950\,mJy\,beam$^{-1}$ in all panels
  the noise is $\sim$1 mJy\,beam$^{-1}$ for the high density tracers, and 
  $\sim$1.5 mJy\,beam$^{-1}$ for the CO at 1.9\arcsec\ resolution.
   The color
  scale is 0--250\,mJy\,beam$^{-1}$ for CO and
  0--22.5\,mJy\,beam$^{-1}$ for the other species with a square root stretch.  The {\em bottom
    right} panel shows the CO spectrum with contours of CN (yellow
  contour at 50\,mJy\,beam$^{-1}$) and HCN (magenta contour
  100\,mJy\,beam${-1}$) overlaid. Note that all data cubes except for
  the CN incorporate total power information from single-dish (Mopra
  for CO, IRAM 30m for the rest, see \S2.1.2).\label{fig:fig5}}
\end{figure*}

From Fig.~\ref{fig:fig4}, we conclude that HCN(1--0) and CO(1--0) emission 
is present along the {\em SW streamer} in the relevant velocity range 
120--250\,km\,s$^{-1}$. In Fig.~\ref{fig:fig5} we show the position velocity diagrams for high--dipole
molecules (HCN, HCO$^+$, CS, and CN) associated with dense gas and
compare them to the CO emission (CN is usually the chemical byproduct
of the photodissociation of HCN). It is clear from Figs.~\ref{fig:fig4} and \ref{fig:fig5} that
the dense gas tracers also show the same outflow characteristics
apparent in CO(1--0), at position offsets $<-5\arcsec$ and
line--of--sight velocities of $<300$\,km\,s$^{-1}$. HCN and HCO$^+$ emission in an outflow
have to date only been reported in a few systems: These include the ULIRG QSO Mrk\,231 (Aalto et
al.\ 2012, Lindberg et al.\ 2016), a system with SFR$\sim$200\,M$_\odot$\,yr$^{-1}$, 70 times
larger than that of NGC\,253. In NGC\,1266, velocity wings are detected in CS(2--1) and HCN(1--0) (Alatalo et al.\ 2015).
In NGC\,1068, Garc{\'{\i}}a-Burillo et al.\ (2014) report dense molecular gas detections in an AGN-driven outflow.
In the starburst galaxy M\,82 the detection of dense gas tracers at the base of its 
molecular outflow (Walter et al.\ 2002, Leroy et al. 2015a) have been reported by Salas et al.\ (2014).

The clear detection of dense gas tracers in the
{\em SW streamer} of NGC\,253 provides an interesting new 
constraint on small spatial scales of
outflowing gas.  Note that the CN emission suffers from artificial
velocity broadening due to blending of hyperfine components: the
brightest CN hyperfine component (1,3/2,5/2 $\rightarrow$ 0,1/2,3/2)
is at 113490.9\,MHz, and the two brightest satellites (a factor of 3
fainter) are (1,3/2,1/2 $\rightarrow$ 0,1/2,1/2) at 113499.6\,MHz and
(1,3/2,3/2 $\rightarrow$ 0,1/2,3/2) at 113508.9 \,MHz, corresponding
to 24\,km\,s$^{-1}$ and 50\,km\,s$^{-1}$ blueward of the main
component (see Skatrud et al.\ 1983 for CN transitions and rest
frequencies).

\begin{figure} \centering
\includegraphics[width=7.0cm,angle=0]{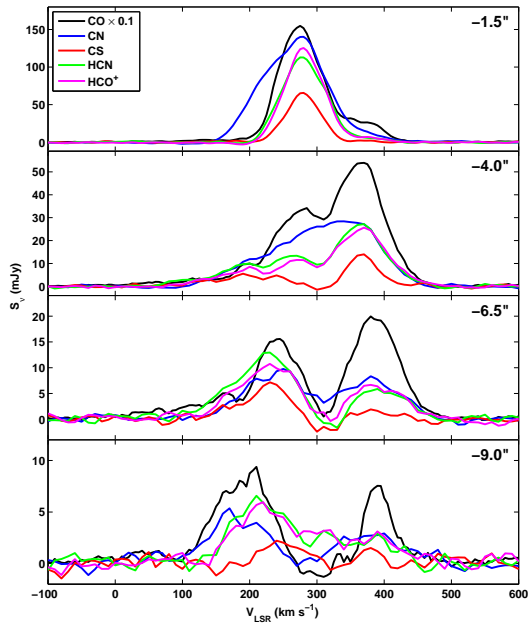} 
\caption{Spectra along the southern outflow at different offsets from
  the centre (shown in the top right of each panel in arcsec, `0'
  offset corresponds to the plane of the disk). The black lines
  indicate the CO emission (scaled down by a factor of 10 for display
  purposes), and the CN, CS, HCN, HCO${+}$ spectra are shown in blue,
  red, green, and magenta, respectively (all data have been convolved
  to the same beam of 1.9$"$). Note that the CN line is artificially
  broadened by the blended hyperfine structure (see \S{3.5}). The
  feature at $\sim$\,200\,km\,s$^{-1}$ is associated with the {\em SW
    streamer}, and the emission at $\sim$\,400\,km\,s$^{-1}$ is from
  the disk. Note how the ratio of CO / (dense gas tracer) is much
  lower in the outflow (and similar there to the central starburst
  region at offsets --1.5$"$) than in the disk, suggesting that dense
  clumps with properties similar to the dense gas in the starburst
  (and of much larger density than gas in typical disk GMCs) survive
  in the outflow. \label{fig:fig6}. }
\end{figure}

We quantify the relation between the different transitions observed in
the streamer using Fig.~\ref{fig:fig6}, in which we present spectra in CO, HCN,
HCO$^{+}$, CS, and CN at different offsets from the center (as before, all data
are on a common resolution of 1.9\arcsec). These spectra show both
emission from the molecular gas streamer (at velocities
$\sim$200\,km\,s$^{-1}$) as well as emission from the disk of NGC\,253
($\sim$400\,km\,s$^{-1}$) which allows us to immediately compare their
line ratios. In the central region we measure peak flux densities for
CO, CN, CS, HCN and HCO$^+$ of 1548.0, 139.8, 66.3, 112.6, and 123.5
mJy~beam$^{-1}$. Consequently, we measure a flux ratio of $\sim13.8$
between CO and HCN, which corresponds to a Rayleigh-Jeans brightness
temperature ratio of $\sim8$. Such ratios, indicating a high dense gas
fraction, is typical of starburst environments (e.g., Gao \& Solomon
2004) and also seen in some of the densest regions of nearby star
forming disks (e.g., Usero et al.\ 2015).

At an offset of $-6.5\arcsec$, significantly off the central starburst
region and clearly corresponding to the kinematically distinct {\em SW
streamer}, we measure the following values for the outflow feature at
$\sim$200\,km\,s$^{-1}$ (in the same order and units as above): 154.0,
9.8, 7.0, 12.8, 10.4. These values yield a CO/HCN peak flux density
ratio $\sim 12$. This is very similar to the value of the CO/HCN ratio
in the starburst, and the same holds true for ratios of CO to the
other molecules. By contrast, the corresponding values for the
emission that we kinematically identify with the background rotating
disk component of the galaxy (visible at a velocity of
$\sim$400\,km\,s$^{-1}$ in Fig. \ref{fig:fig6}) are 193, 8.4, 1.9, 5.6, 6.4 (same
order and units as above). These yield a CO/HCN peak flux density
ratio $\sim35$, a factor of 3 higher than the starburst or the
streamer. Such a value for the CO/HCN ratio is in excellent agreement
with typical measurements in the {\em disks} of nearby galaxies (Usero
et al.\ 2015). {\em We conclude that the fraction of dense molecular
gas in the {\em SW streamer} of the outflow is
high and stays high throughout the region where we detect emission.}
Moreover, the  material in the molecular streamer displays a
dense gas fraction that is similar to that of the starburst region and
significantly higher than that of the main disk of NGC\,253, strongly
suggesting a starburst origin for the gas in the streamer.

\subsection{Mass and Density of the SW Streamer}

A key result in the study of Meier
et al.\ (2015) is that the CO(1--0), HCN(1--0), and HCO$^{+}$(1--0)
transitions are optically thick (and in fact have similar optical
depths, $\tau\sim5$) in the central region of NGC\,253. The fact that
the line ratios we measure do not change from those in the central
region suggests that their emission is also optically thick in the
{\em SW streamer}. This has important potential consequences for the
molecular gas mass in the outflow. The CO(1--0) luminosity of the
streamer is $L_{\rm CO}\sim2.8\times10^6$\,K\,km\,s$^{-1}$\,pc$^2$,
with an approximate area of $\sim1.1\times10^4$\,pc$^2$ and an
integrated surface brightness of $T_{\rm CO}\Delta
v\sim250$\,K\,km\,s$^{-1}$. In order to obtain a firm lower limit for
its mass Bolatto et al.\ (2013) used a CO-to-H$_2$ conversion factor
$\alpha_{\rm CO}=0.34$\,M$_\odot$\,(K\,km\,s$^{-1}$\,pc$^2$)$^{-1}$
based on optically thin CO(1--0) emission. This leads to a ``minimum
mass'' of the streamer of $\sim$10$^{6}$\,M$_\odot$ ($\sim13$ times
lower than what one would derive with the typically assumed Galactic
conversion factor).

\subsubsection{ Inferred volume densities and implications for dense gas
  excitation}

The projected area of the streamer is $\sim$240$\times$60\,pc$^2$ (or
$\sim10^{41}$\,cm$^2$), leading to an average H$_2$ column density of
$N({\rm H_2})\sim4\times10^{21}$\,cm$^{-2}$ in the plane of the
sky. If we assume that the feature has cylindrical geometry
(depth\,$\sim$\,width) this results in an average volume density of
40\,cm$^{-3}$. This is completely insufficient to excite the detected
dense gas tracers through collisions: for example, the critical
density (defined as the density at which the rate of the collisional
depopulation of a quantum level equals the spontaneous radiative decay
rate) for the HCN(1--0) emission is 2.6$\times10^6$\,cm$^{-3}$. As is
frequently pointed out, however, what is more relevant to the
excitation is the `effective excitation density', which includes the
summation of all collisional transitions to the lower level and the
effect of radiative trapping in an optically thick environment
(e.g., Stacey et al.\ 2011, Shirley et al.\ 2015). The effective
excitation density can be 1--2 orders of magnitude lower than the
critical density even for HCN, particularly in environments of high
optical depth (Shirley et al.\ 2015). Nonetheless, even after taking
these effects into account our derived average gas density would be
still 2--3 orders of magnitudes too low to explain the presence of
bright emission from HCN and other high-dipole molecules. Note also
that the line-widths are large (Fig. \ref{fig:fig6}), which lowers the
effect of radiative trapping.

There are a number of factors that could bring these numbers in
agreement. First of all is the clumping -- the molecular gas is likely
not distributed smoothly. The observed surface brightness in the {\em SW
streamer} at 32~pc resolution is T$_{\rm b}^{obs}\sim1.5-2$~K. Assuming an
intrinsic brightness of T$_{\rm b}\sim100$~K (not unreasonable since
the gas in the streamer is likely warm), the resulting area clumping factor
would be $\sim$T$_{\rm b}/$T$_{\rm b}^{obs}\approx50-70$, corresponding to a clumping factor in
volume of $50^{3/2}\sim350-600$. This implies that our bulk density of
$40$\,cm$^{-3}$ is physically closer to $2\times10^4$\,cm$^{-3}$, with
typical column densities of N(H$_2$)$\sim10^{22}$\,cm$^{-2}$. Note
that these estimates come from the ``minimum mass'' yielded by the
assumption of an optically thin CO--to--H$_2$ conversion
factor. Depending on the optical depth of the CO(1--0) transition, the
actual conversion factor (and consequently the derived molecular mass
and resulting mass outflow rate) could be factors of few to several
times higher. To produce bright HCN emission it would be more
comfortable to raise the mass in the streamer by a factor of a few,
which would raise the physical volume density to $10^5$\,cm$^{-3}$ and
reach conditions where HCN(1--0) would be efficiently excited.
Following this line of reasoning, the bright HCN emission in the
streamer suggests that the minimum molecular outflow rate of
3\,M$_\odot$\,yr$^{-1}$ is probably an underestimate by factors of a
few (Bolatto et al.\ 2013).

\subsubsection{Excitation by electrons?}

Collisions with electrons are also an  effective
way of exciting polar molecules (Dickinson et al.\ 1977). In fact they
are the main mechanism for HCN excitation in comets (Lovell et
al.\ 2004).  Calculations show that the rate coefficient for
excitation of HCN(1--0) by collisions with electrons is
$>10^5$ times larger than for collisions with H$_2$ (Saha et al.\ 1981;
Faure et al.\ 2007).  The reason why electrons are usually not
considered as important collisional partners for HCN is that the
electron fraction, $x_e$, in dark molecular clouds is thought to be
$x_e\sim 10^{-8}$ or lower, controlled by cosmic ray ionization (e.g.,
Flower et al.\ 2007).  But for $x_e\gtrsim10^{-6}$ excitation by
collisions with electrons will be important and likely dominant. The
{\em SW streamer} is close to and probably embedded in the ionized
flow, so it is possible that it possesses a higher ionization fraction
than a typical dark cloud, and HCN(1--0) is mostly excited by
electrons. This could also be the case if, for example, the outflow is
cosmic-ray driven. We note, however, that the line ratios in the {\em
  SW streamer} are very similar to the line ratios observed in the
starburst (Fig. \ref{fig:fig6}), which makes it unlikely that the excitation 
mechanisms that dominate in one region do not also dominate in the other.

\subsubsection{Constraints to mass and surface density from extinction}

We can use the extinction inferred for the {\em SW streamer} from the
observed H$\alpha$ to Pa$\beta$ line ratio to independently
constrain its molecular mass.  The typical H$\alpha$/Pa$\beta$ ratio
in the streamer is $R_{obs}=4.5-5$, while the expected intrinsic line
ratio in a 10$^4$~K plasma is $R_{int}\approx17.5$ for case
B~recombination (Osterbrock \& Ferland 2006). For a Cardelli, Clayton,
\& Mathis (1989) extinction curve, the ratio of extinction at
1.28\,\micron\ to extinction at 0.66\,\micron\ is $r_{P\beta
  H\alpha}=A_{\rm Pa\beta}/A_{\rm H\alpha}\simeq1/3$.  Thus explaining
the observed line ratio with a single screen of extinction in front of
the ionized gas requires $A_{\rm
  H\alpha}=-2.5\log\left[R_{obs}/R_{int}\right]/(1-r_{P\beta
  H\alpha})\approx2-2.2$, which results in $A_V\sim2.5$. For the Milky
Way, this implies a molecular column density in the plane of the sky
of $N(H_2)\sim 2.5\times10^{21}$\,cm$^{-2}$ (Bohlin, Savage, \& Drake
1978). The screen geometry is a lower limit to the column density,
since any unextincted ionized gas emission in front of the screen
leads to an increase in $R_{obs}$, while conversely any extinction on
the back side of the ionized emission would remain undetected. To
account for some of these effects a ``double screen'' geometry is
usually preferred, leading to $N(H_2)\sim5\times10^{21}$\,cm$^{-2}$.
We would reach a 20\% larger higher value if we assumed that the dust
was mixed with the gas. These numbers are consistent with the ``lower 
limit'' mass estimate from CO inferred
above. Optically thin CO emission, however, is somewhat surprising
given the relatively low CO/HCN ratio in the streamer, reminiscent of
the central starburst region.  Thus it is likely that there is
significant ionized gas in front of the streamer.  ALMA observations
of higher--J CO lines will help to constrain the properties of the
{\em SW streamer} and the conditions in the molecular outflow further.

\subsection{The driving of the SW Streamer}

The mechanisms to impart momentum and accelerate cold gas in a galactic
outflow are not well understood.  The presence of molecular material
$\gtrsim$300\,pc away from the central bar, particularly in the form
of dense gas as traced by HCN, HCO$^+$ and CS, already provides strong
constraints on the stability of wind-entrained clouds against photo
and thermal evaporation, Kelvin--Helmoltz instabilities, and shedding
due to ablation (e.g., Marcolini et al.\ 2005). Simulations of
radiative clouds embedded in a supersonic flow show that radiative
cooling stabilizes clouds against destruction (Cooper et
al. 2009). Radiative clouds undergo a lower acceleration and have a
higher Mach number relative to the flow than adiabatic (energy--conserving) clouds. They do
experience fragmentation due to Kelvin--Helmholtz instability, but the
resulting cloudlets are denser and are more resistent to destruction
than adiabatic clouds, and they are drawn into the flow creating
filaments. Additionally, magnetized clouds entrained in a hot wind
have also been shown to be more stable than unmagnetized ones, even
for very moderate initial internal magnetic fields, while also leading
to filamentary structures (McCourt et al.\ 2015). In a set of recent
calculations, Scannapieco \& Br\"uggen (2015) show that supersonic
flows suppress the Kelvin-Helmholtz instability thus allowing cooling
clouds to survive longer, but the compression of the cloud also makes
its acceleration less efficient so that they only reach $\sim15\%$ of
the hot flow velocity before being disrupted.  The inclusion of cloud
evaporation induces further compression and naturally produces highly
elongated, filamentary clouds similar to the streamers we observe, but
although it can help the clouds survive even longer it also makes the
momentum transfer less efficient, resulting in even lower final
velocities before disruption (Br\"uggen \& Scannapieco 2016). Very recent simulations
by McCourt et al.\ (2016), however, suggest that crushed clouds shatter
into tiny, dense, cloudlets that do no disappear and can be much more easily 
accelerated by the hot wind, making the entrainment process efficient. In this scenario 
the {\em SW streamer} could be composed of a collection of dense molecular 
clumps embedded in the much hotter outflow.

Notably, a feature corresponding to the molecular gas streamer is
also seen in radio continuum maps obtained at 20\,cm (Fig.~9 in
Ulvestad et al.\ 1997, see also Heesen et al.\ 2009), implying that
synchrotron emission is associated with the outflowing molecular gas.
This suggests that magnetic fields, and possibly cosmic rays, are
associated with the outflow. Cosmic rays have been suggested as an
important mechanism for driving outflows in galaxies (e.g., Booth et
al.\ 2013, Salem \& Bryan 2014, Girichidis et al.\ 2016, Simpson et al.\ 2016), although
observationally constraining their importance remains elusive.

To impart momentum to the cold gas, an alternative to
entrainment in a hot flow is acceleration by radiation pressure. In order to explore
the viability of radiation pressure to explain the observations, we
perform a rough calculation. The radiation flux at a distance $r$ along its axis from
the center of a uniform-brightness disk of radius $R_{src}$ and
luminosity $L_{src}$ is $F=L_{src}/[2\pi(R_{src}^2+r^2)]$.  Only the photons 
absorbed by the dusty cold gas can impart
momentum, so the force due to radiation pressure experienced by a
cloud that subtends a solid angle $\Omega$ and has an optical depth $\tau$ is
$f_{rad} = F\,\Omega\,r^2 (1-e^{-\tau})/c$.  The resulting velocity is
then simply $v_{cloud}^2=2\int\,f_{rad}/M_{cloud}\,dr$. For the
purposes of the calculations below we will assume that the {\em SW
  streamer} is exposed to radiation from the ``naked'' starburst,
which would be mostly far-ultraviolet and easily absorbed
($\tau\gg1$).  There is the potential for the momentum imparted by the radiation
to be boosted by a factor of order $\tau_{\rm IR}$, the mean infrared optical depth of the streamer
(e.g., Thompson et al.\ 2015; Zhang \& Davis 2016). Given our column density estimates, 
the {\em SW streamer} does not have a large enough column to possess substantial 
optical depth to its own reradiated IR photons ($\tau_{\rm IR}\sim 0.2$ for the most 
favorable geometry), so we assume that this boosting is unimportant.

NGC\,253 has a total infrared luminosity of L$_{\rm
  TIR}=1.4\times10^{44}$\,erg\,s$^{-1}$, about half of which
originates within a diameter of $\sim200$\,pc from its center (Leroy
et al.\ 2015), i.e. $R_{src}$=100\,pc. The {\em SW streamer} has a ``minimum mass''
$M_{cloud}\approx10^6$\,M$_\odot$, a projected size on the sky of
$\sim60\times240$\,pc$^{2}$, and a typical distance to the center of
order $r\sim100$\,pc. If we assume the geometry corresponds to a
cylinder 60\,pc in diameter ($\Omega r^2=\pi (30\,{\rm pc})^2$) illuminated by
$L_{src}=10^{44}$\,erg\,s$^{-1}$, the force due to radiation pressure
applied at the base of the streamer would be
$f_{rad}\sim7.5\times10^{31}$ dyn, resulting in an estimated
acceleration of $a_{rad}\sim4\times10^{-8}$\,cm\,s$^{-2}$. In the
absence of gravity, the resulting velocity at $r\sim100$\,pc would be
$v_{cloud}\sim60$\,km\,s$^{-1}$, lower than the observed velocity
before projection correction (Fig. \ref{fig:fig3}). If we instead
assume that the radiation is exerting pressure on the long side of the
cylinder (the geometry with the maximum area), $f_{rad}$ would be 5
times larger, with a corresponding
$v_{cloud}\sim135$\,km\,s$^{-1}$. Note, however, that the ratio of
gravity to radiation pressure for stars is
$f_{grav}/f_{rad}=G\,\Sigma_{cloud}\,(4\pi c)/\Psi$, where
$\psi\approx2000$\,erg\,s$^{-1}$\,g for a fully sampled IMF containing
young stars, and $\Sigma_{cloud}$ is the surface density of the
cloud ($\Psi$ is the light--to--mass ratio of the stellar population,
e.g., equation 33 in Kim et al.\ 2016). For the end-on geometry
$\Sigma_{cloud}\sim0.075$\,g\,cm$^{-2}$, while for the side-on
geometry $\Sigma_{cloud}\sim0.015$\,g\,cm$^{-2}$, resulting in
$0.94\geq f_{grav}/f_{rad}\geq 0.19$. So for our assumed geometry and
mass the net acceleration $a_{net}=a_{rad}(1-a_{grav}/a_{rad})$ is
between 6\% and 81\% of $a_{rad}$, which would result in $15\leq
v_{cloud}\leq 120$\,km\,s$^{-1}$.

From the above estimates it appears that radiation can contribute maybe
up to a few tens of percent of the momentum, but it is extremely unlikely to
completely explain the observations. Doing so would require using the
most favorable geometric assumption (maximal cloud area, which
maximizes $f_{rad}$ and minimizes $f_{grav}/f_{rad}$) and also
assuming that there is almost no projection correction to the measured
velocity (which requires a contrived geometry). This suggests that
radiation pressure is not the dominant mechanism for accelerating the
streamer, although it is non-negligible. Note also that any increase
in the mass of the streamer over its minimum mass (see discussion in
\S3.6.1), or accounting for the old stars in the galaxy disk outside
the starburst (which would increase gravity), would also reduce the
importance of radiation pressure.  Much higher resolution observations
of the {\em SW streamer} may help further elucidate these questions,
in particular which mechanisms are driving the molecular gas in the
outflow.

\section{Summary}

Our new ALMA band 3 observations (CO, and dense gas tracers) of the
central starburst region of NGC\,253 give new insights on the
properties of the molecular outflowing gas originally discussed in
Bolatto et al.\ (2013). The most prominent outflow feature towards the
south, which is the main subject of this paper, possesses a large
intrinsic velocity dispersion. Its extent and dispersion are
consistent with an ejection from the disk starting about $\sim1$\,Myr
ago.

It is currently unclear whether or not the molecular mass entrained in
the outflow will escape the galaxy, or be recycled to fuel later
episodes of star formation. For most of the gas in the observed
outflow to escape the galaxy it would need to be accelerated as it
moves away from the disk. The kinematics of the molecular gas are
consistent with accelerating with a velocity gradient of
1\,km\,s$^{-1}$\,pc$^{-1}$, and at its last measurable point it
approaches the escape velocity. As discussed, this
interpretation is not unique: the kinematics could be also consistent
with an outflow with a range of speeds, where the material farther
from the disk is there because it is the fastest. In that scenario,
only the fastest fraction of the outflowing gas may escape the
galaxy. Approved, more sensitive ALMA observations will trace the
outflow even further out, and will shed light on whether or not part
of the ejected molecular material will escape the galaxy.

Strikingly, tracers of the dense gas phase of the molecular medium
(HCN, HCO$^+$, CS, CN) are also spatially coincident with the {\em SW
  streamer} of the molecular outflow. The line ratios HCN/CO of
$\sim1/10$ measured in the outflow are high and consistent with ratios
observed in the central starburst region of NGC\,253 and in other
starbursts. The HCN/CO line ratio in the {\em disk}, on the other
hand, is significantly lower ($\sim1/30$), typical of gas in the disks
of nearby galaxies. In principle, this is indicative the dense
molecular gas being ejected from the central regions into the outflow,
while retaining its properties in this process. It also suggests that
the CO(1--0) and HCN(1--0) emission are optically thick also in the
streamer, implying that the estimated mass loading parameter
$\eta\gtrsim3$ (Bolatto et al.\ 2013) is likely a lower limit. Note
that electron excitation of polar molecules is an often-ignored
mechanism that could play an important role at exciting HCN, HCO$^+$,
and CS emission in outflows, although the fact that we see the same
ratios of these transitions to CO(1--0) in the starburst and the
streamer suggests similar excitation mechanisms in both regions,
implying dense gas is the most likely cause. Simple calculations indicate
that radiation pressure is not the main mechanism for driving the outflow.
The presence of a dense gas phase in molecular outflows (with volume densities
$>$10$^4$\,cm$^{-3}$ and probably $\sim10^5$\,cm$^{-3}$) will have to
be accounted for in numerical simulations of galactic winds, both at
low and high redshift.

\acknowledgements

We thank the referee for excellent comments that improved the paper.
D.S.M. acknowledges partial support by the National Science Foundation
through grant AST-1009620. S.V. acknowledges NSF grant
AST-1009583. A.D.B. acknowledges visiting support by the Alexander von
Humboldt Foundation, and support by the National Science Foundation
through a CAREER grant AST-0955836 and AST-1412419. E.C.O. is
supported by the National Science Foundation through grant
AST-1312006. This paper makes use of the following ALMA data:
ADS/JAO.ALMA \#2011.0.00172.S, \#2012.1.00108.S. ALMA is a partnership
of ESO (representing its member states), NSF (USA), and NINS (Japan),
together with NRC (Canada), and NSC and ASIAA (Taiwan), in cooperation
with the Republic of Chile. The Joint ALMA Observatory is operated by
ESO, AUI/ NRAO and NAOJ. The NRAO is a facility of the National
Science Foundation operated under cooperative agreement by Associated
Universities, Inc. Based on observations made with the NASA/ESA {\em
  Hubble Space Telescope}, obtained from the data archive at the Space
Telescope Science Institute (STScI). Some of the HST data presented in
this paper in this paper were obtained from the Mikulski Archive for
Space Telescopes (MAST), others where acquired under program
HST-GO-13730 with support provided by NASA through a grant from the
STScI. STScI is operated by the Association of Universities for
Research in Astronomy, Inc., under NASA contract NAS5-26555. The Mopra
radio telescope is part of the Australia Telescope National Facility
which is funded by the Australian Government for operation as a
National Facility managed by CSIRO. Based on observations carried out
under project number 209-14 with the IRAM 30m Telescope. IRAM is
supported by INSU/CNRS (France), MPG (Germany) and IGN (Spain).

\end{document}